\documentclass[%
 aip,
 amsmath,amssymb,
 reprint,%
]{revtex4-1}

\usepackage{graphicx}
\usepackage{dcolumn}
\usepackage{bm}

\usepackage[utf8]{inputenc}
\usepackage[T1]{fontenc}
\usepackage{mathptmx}
\usepackage{etoolbox}

\usepackage[dvipsnames]{xcolor}

\makeatletter
\def\@email#1#2{%
 \endgroup
 \patchcmd{\titleblock@produce}
  {\frontmatter@RRAPformat}
  {\frontmatter@RRAPformat{\produce@RRAP{*#1\href{mailto:#2}{#2}}}\frontmatter@RRAPformat}
  {}{}
}%
\makeatother
\begin{document}

\preprint{AIP/123-QED}

\title[Unveiling the bi-stable character of stealthy hydrogen-air flames]{Unveiling the bi-stable character of stealthy hydrogen-air flames}

\author{Rub\'en Palomeque-Santiago}
\affiliation{Dpto.~de Ing. T\'ermica y de Fluidos, Universidad Carlos III de Madrid, 28911, Legan\'es, Madrid, Espa\~na}
\author{Alba Domínguez-González}
\affiliation{ETSIAE., Universidad Polit\'ecnica de Madrid, Plaza del Cardenal Cisneros 3, 28040, Madrid, Espa\~na}
\author{Daniel Mart\'inez-Ruiz} 
\affiliation{ETSIAE., Universidad Polit\'ecnica de Madrid, Plaza del Cardenal Cisneros 3, 28040, Madrid, Espa\~na}
\author{Mariano Rubio-Rubio}
\author{Eduardo Fern\'andez-Tarrazo} 
\author{Mario S\'anchez-Sanz$^*$}
  \email{mssanz@ing.uc3m.es}
\affiliation{Dpto.~de Ing. T\'ermica y de Fluidos, Universidad Carlos III de Madrid, 28911, Legan\'es, Madrid, Espa\~na}

\date{\today}

\begin{abstract}

Ultra-lean hydrogen-air flames propagating in narrow gaps, under the influence of cold walls and high preferential diffusion, can form two distinct isolated structures. They exhibit either circular or double-cell shapes and propagate at different speeds, with the latter roughly doubling in size and traveling speed to the former. 
Hydrogen mass diffusivity, convective effects and conductive heat losses are the physical mechanisms that explain the alterations in morphology and propagation speed. 
In previous experiments, Veiga et al.~Phys.~Rev.~Lett.~124, 174501 (2020) found these clearly distinguished flame structures for different combinations of equivalence ratio, channel gap and the effect of gravity on the dynamics of upwards and downwards propagating flames in a vertical chamber. Present observations in horizontal channels show the simultaneous appearance of these two stable structures, which arise under identical experimental conditions and conform the first evidence that multiplicity of stable solutions coexists in real devices. To explain the observations, we performed numerical simulations using the simplified model, which show that symmetry-breaking details during the ignition transient explain the concurrent emergence of the two stable configurations.  
This discovery urges the need to implement additional engineering tools to account for the possibility of formation and propagation of isolated flame kernels at different speeds in hydrogen-fueled systems.
\end{abstract}

\maketitle

\section{Introduction}
Molecular hydrogen (H$_2$) is recognized as a crucial energy carrier for achieving Net Zero greenhouse emission goals in the coming decades. The anticipated increase in its utilization in the power generation and transport sectors has raised legitimate concerns about associated risks and threats \cite{staffell2019role, SALVI20151132}. Understanding the initiation and progression of chemical reactions in gaseous mixtures is vital for advancing safe hydrogen-powered devices. Among the possible scenarios, we consider here the leaking of hydrogen from a container that, after mixing with the surrounding air, can form a reactive mixture. If this mixture comes into contact with an energy source, it might ignite to form a potentially dangerous traveling premixed flame. The narrow channel configuration addressed here is a confined space that serves as representation of fuel leakage accidents in caged devices.
The growing use of hydrogen in the future and its singular properties as a fuel (wide flammability limit, low ignition energy and almost invisible flame) will certainly increase the risks during its utilization. 

\begin{figure*}
\centering
\includegraphics[width=0.75\textwidth]{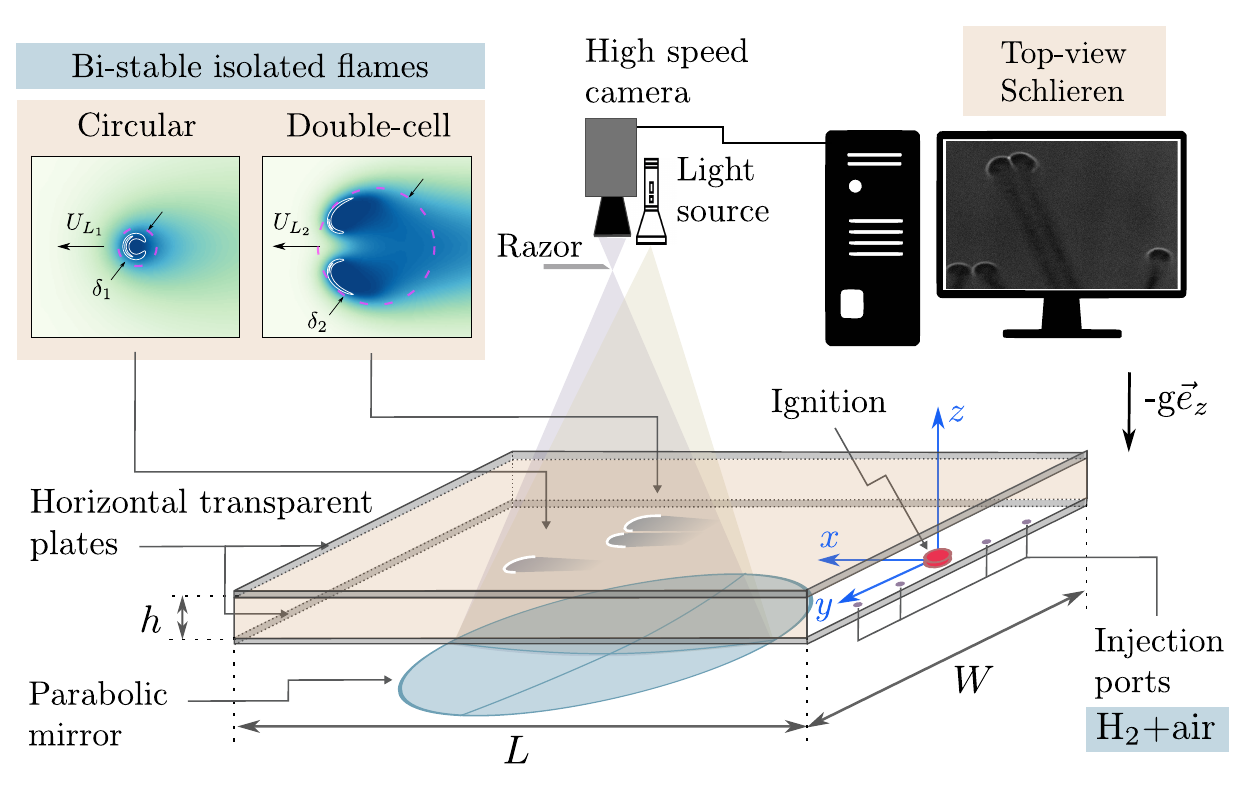}
\caption{Schematic of the horizontal chamber and experimental system setup used for image acquisition. The dimensions of the gap are $L \times W \times h$, with $L=950$ mm,  $W=220$ mm and $ 2<h<6$ mm. }
\label{fig:expsetup}
\end{figure*}

The thermo-chemical characterization of this type of flames is typically addressed from the amenable and simplified planar flame (1D) theory \citep{sanchez2014recent}, a useful starting point but insufficient to explain experimental observations. 
These deviations from the theoretical framework are particularly pronounced as the H$_2$ concentration in air decreases. In fact, the planar flame construct does not exist for the ultra-lean conditions studied here,  when preferential diffusion effects take part in enabling the survival of curved reactive fronts that display locally richer and hotter conditions. 

Stable multidimensional isolated flame structures, such as three-dimensional flame balls, have been predicted theoretically and observed experimentally \citep{ronney1994stationary}. On the other hand, their steady two-dimensional counterpart had been previously considered nonexistent in theoretical analyses \citep{Matkwosky1979}, and they have only been found as drifting flame cells in two-dimensional simulations \citep{grcar2009new}. 

However, Veiga-López et al.\cite{veiga2020unexpected} found   that various types of nearly two-dimensional stable isolated-flame configurations can be found under confinement due to  geometrical constraints and additional viscous and conductive processes. They found multiple quasi two-dimensional propagating fronts in lean hydrogen-air mixtures in a vertical Hele-Shaw cell. Depending on the controlling parameters, channel width and mixture composition, the flames remained stable propagating at constant speed or divided forming a fractal-like path. These isolated flame fronts, similar to flame balls and drifting flames, are  characterized by their small size and low emissivity, which makes them very elusive. Among the atlas of behaviors, only two stable cases were detected. Based on the equivalence ratio of the mixture and the flame propagation direction (upwards or downwards) either circular or double-cell isolated traveling flames were produced, but never observed simultaneously. Circular flames burn slowly and were observed during upward propagation. Double-cell structures travel faster and were only observed during downward propagation. It was thus concluded that buoyancy effects would determine the prevalence of one type over the other by either accelerating or decelerating the front owing to the placement of hot products. This result confirmed the conclusions of previous numerical studies that included buoyancy effects \cite{Martinez2019}.    

An exhaustive parametric study  by Domínguez-González et al.\citep{dominguez2022stable} detailed the occurrence of  numerically stable solutions in the nearly 2D limit with heat losses. Therein, it was shown that both configurations, circular and double, can arise in a common region of the parametric space subject to different initial conditions. It is remarkable that given a prescribed mixture and vessel properties, two stable flames, of different sizes and propagation speeds, can be formed. 

To extend these previous studies\citep{veiga2020unexpected,dominguez2022stable}, we experimentally address the formation of both isolated flame structures in real devices. To do so, we designed and constructed a new horizontal experimental set-up to eliminate the alignment of buoyancy effects with the propagation direction. This new experimental campaign shows that the two stable flame structures can coexist under identical geometric and mixture conditions. Such simultaneous coexistence of two stable canonical configurations, which has not been observed before, constitutes the core of this manuscript and represents a unique case in the literature, to the best of the authors' knowledge.

The paper is organized as follows. We first offer a description of the configuration at hand and define the characteristic scales of the problem. The detailed presentation of the experimental setup and data processing methods comes next. Along the paper we identify the parameters that control the onset of bistability in these flames to compare our results with recent numerical studies. Finally, we discuss the experimental and numerical results before presenting our conclusions.

\section{Nearly two-dimensional configuration}
We shall consider in the following a horizontal Hele-Shaw chamber, formed by two parallel plates that are separated a small distance apart $h$, in the millimetre range, as illustrated in figure~\ref{fig:expsetup}. This particular configuration serves as a representative model for confined geometries and enables the simplification of the analysis by reducing one of the spatial dimensions to the constrained length $h$, perpendicular to the propagation direction. The narrow gap between the plates results in a reduced combustion volume, creating a competition between heat generation at the flame and conductive heat losses through the walls.
These cases are of wide interest in combustion science \citep{ballossier2024three, martinez2019premixed}, being typically easier to control, measure and analyze than 3D experiments of premixed flames \citep{ronney1998experimental} in absence of wall effects.

Although extinction and quenching can be anticipated for an extreme reduction of the channel gap ($h\lesssim 1$~mm), it is possible to find stationary flames in the combined regime of ultra-lean hydrogen-air mixtures with a slightly larger plate separation\cite{veiga2020unexpected}.  
Considering that the isolated cell flame is of characteristic size $\delta_i$ and it propagates with velocity $U_{L_i}$, we can estimate of the heat flux from the flame to the horizontal plates as $q_k \sim  \delta_i^2 k_g (T_a - T _u)/h$. The subindex $i$ identifies a circular flame when $i=1$ and a double-cell flame when $i=2$. The factor  $\delta_i^2$ represents the effective hot area in contact with the plates, $k_g$ is the gas conductivity, $T_a$ is the adiabatic flame temperature and the plates are considered to be a heat sink at room temperature $T_u$. 

On the other hand, the heat released by the flame per unit time can be estimated as $q_f \sim \rho U_{L_i} c_p (T_a-T_u) h \delta_i$, where $c_p$ and $\rho$ represent the heat capacity and density of the gas, respectively, and $h\delta_i$ is the approximate flame surface. The comparison between the heat losses to the plates and the heat released by the flame yields the relative production-loss parameter 
\begin{equation} 
\Delta=\dfrac{q_k}{q_f}= \dfrac{\delta_i}{h} \dfrac{\delta_T}{h} \label{eq:Delta},
\end{equation}
which must remain below unity.
In this expression, the thermal thickness of the flame is $\delta_T \sim D_T/U_{L_i}$, where $D_T = k_g/(\rho c_p)$ is the thermal diffusivity of the gaseous mixture. Moreover, the value $U_{L_i}$ is used to define $\delta_T$ instead of the planar flame propagation speed $S_L$, traditionally chosen to theoretically estimate the thermal flame thickness. In the ultra-lean mixture regime tested in our experiments, always with equivalence ratio below $\phi< 0.3$,  planar flames do not exist\citep{sanchez2014recent} and, therefore, $S_L$ is not a representative value of the flame propagation velocity. 

It will be shown below (figure~\ref{fig:bifurcation}) that the characteristic size of the flame kernel is similar to the gap distance $\delta_i \sim h$. Considering this, it is easy to see from equation~\eqref{eq:Delta} that a progressive reduction of the distance $h$ will increases the relative relevance of the conductive heat losses through the plates with respect to heat production. Therefore, it is expected that the corrugated continuous flame fronts will split forming isolated weak reaction kernels as $h$ is reduced. This point has been demonstrated in recent studies, both numerical~\citep{Martinez2019} and experimentally~\citep{veiga2020unexpected}, in vertical narrow channels with H$_2$ concentrations as low as 5\%~vol.  
 
The characteristic size of the isolated flames $\delta_i$ is determined by the most unstable wavelength of the thermodiffusive instability \citep{clavin2016combustion} and by the preservation of activated chemical kinetics through curvature. The latter, in turn, relies on the preferential diffusion of hydrogen-air mixtures, where a large mass-to-thermal diffusivity ratio prescribes low values of the Lewis number ${Le}=D_{T}/D_{i}=0.3$.  {\color{black} The onset of thermodiffusive} instability accelerates the wrinkling process of the flame front, forming regions with positive and negative curvature\cite{Martinez2019}. Moreover, the aforementioned heat losses induce major inhibition of the reaction at the concave part of the wrinkled front, inducing partial extinction and dividing the reactive front into different kernels that propagate steadily, leaving a significant portion of the mixture unburned. {\color{black} Without the effect of thermodiffusive instabilities, minor heat losses would quench the flame, as was demonstrated by Martínez-Ruiz et al.\cite{Martinez2019} in their calculations with $Le=1$}. An example of the two different stable isolated fronts that are formed, either circular (slower propagation at speed $U_{L_1}$) or double-cell (faster propagation at speed $U_{L_2}$), can be seen in the inset of figure~\ref{fig:expsetup}. These flames benefit from conductive heat losses through the confining plates to enable their steady propagation solutions, in a similar manner to flame balls or drifting-comet flames requiring radiation losses \cite{zeldovich1944theory} or convective effects \cite{grcar2009new}, respectively. 

An attempt to study this problem using direct three-dimensional simulations is presented in the work by Melguizo-Gavilanes et al.\cite{melguizo2021three}. These simulations, that used detailed chemistry and transport, validated the quasi-2D approximation for values $\Delta \ll 1$ in the adiabatic case. The calculations encompass both the flow field and the thin flame, where the chemical reaction occurs. Typically, the flow field is hundreds or thousands of times larger than the flame, and the grid in both regions must be  sufficiently fine to numerically integrate the equations accurately. Such disparity between the characteristics lengths scales quickly escalates the problem to reach tens of millions of unknowns. Consequently, conducting a parametric study to assess the impact of heat losses and buoyancy becomes prohibitively expensive, requiring hundreds of thousands of computational hours \citep{LU2009192}, even when employing simplified chemical kinetics models.

The calculation cost can be eased by performing an asymptotic analysis in the narrow-channel limit approximation. This enables an adequate separation of scales valid when the characteristic flow length scale $\ell > \delta_i$ and plate size $L$ are larger than the gap separation $h$. The associated residence time at scales of the order of the structure itself is comparable to the heating time of the horizontal plates $t_{r} \sim \delta_i/U_{L_i} \sim t_h \sim h_w^2/D_{T_w}$. Therefore, an effective heat transfer to the plates is achieved during the flame propagation inside the channel. In this limit, a linear variation of temperature across the thickness of the walls can be considered \citep{Martinez2019}.

All things considered, the 3D conservation equations can be substituted by a quasi-two-dimensional viscosity-controlled formulation with conductive heat losses\citep{Martinez2019}, which was recently used to delineate the parametric space of multiple stable isolated solutions by Domínguez-González et al.\citep{dominguez2022stable}. 
These results were later expanded by Domínguez-González et al.\citep{dominguez2022stable}. In their simulations, they found stable isolated circular and double-cell flames with both positive and negative buoyancy by tweaking the initial conditions, a result that seems to broaden the experimental observations by Veiga-López et al.\citep{veiga2020unexpected}. 
 
\begin{figure}
\centering
\includegraphics[width=\linewidth]{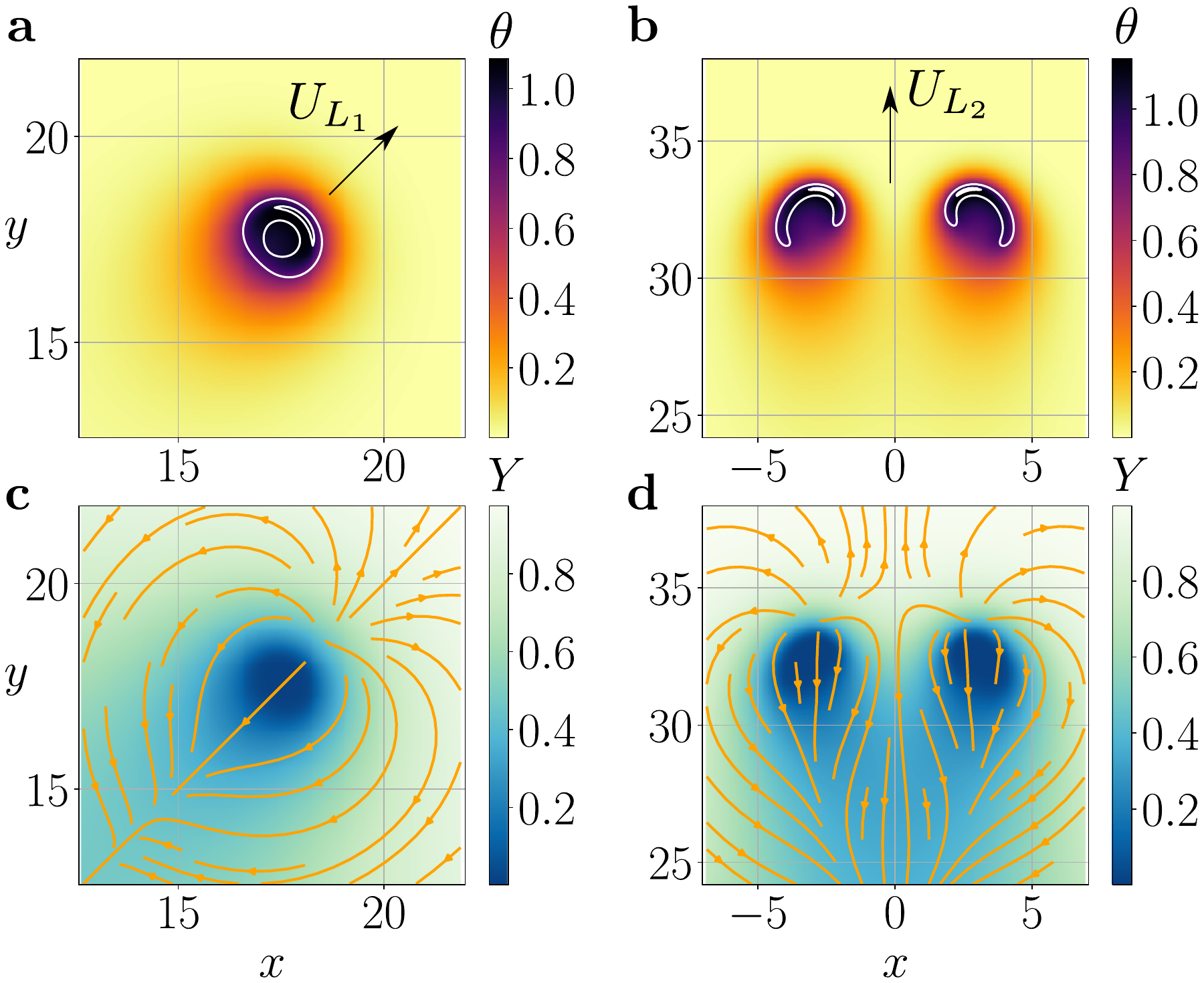}
\caption{Simulations of stable isolated fronts for the same parametric set $q=5$, $b=1$. Steady-state dimensionless temperature $\theta = (T-T_u)/(T_a -T_u) $ and reactant concentration $Y$ are shown for a circular flame (\textbf{a}, \textbf{c}) and a double-cell flame (\textbf{b}, \textbf{d}). Isocontours of reaction rate (white) and flow streamlines (orange) are included in top and bottom panels, respectively.
}
\label{fig:simul}
\end{figure}

In this work, we performed new transient gravity-free simulations using the quasi-two-dimensional pseudo-spectral (Fourier-Fourier) code developed by Domínguez-González et al.\citep{dominguez2024numerical}, to obtain the two stable configurations shown in figure~\ref{fig:simul}. A comparison of the ignition transient properties with the experimental measurements will be shown below in section~\ref{sec:ignition}. 
In particular, values of the dimensionless heat release ${q=({T_a - T_u})/{T_u}= [2-5]}$ are chosen such that the adiabatic temperature reaches ${T_a = (1+q) T_u \simeq [800-1600]}$~K, a characteristic range of ultra-lean equivalence ratios ${\phi \simeq [0.2-0.35]}$ under atmospheric conditions \citep{fernandez2009one}. Note that the flame front needs to remain curved to reach superadiabatic temperatures that stay above the crossover value, an effect that is only observed in low-Lewis number fuels like hydrogen. This crossover temperature is around 1000 K at atmospheric pressure \citep{sanchez2014recent} and is defined as the temperature at which the rates of the branching and recombination steps in the chemical mechanism are equal.

The numerical results presented here make use of dimensionless variables, where $\tau=t/t_r$ is the dimensionless time referred to  $t_r = (D_{T}/U_{L_i}^{2})$ and $x$, $y$ are dimensionless coordinates referred to $\delta_{T}$. The transverse-averaged energy conservation equation
\begin{equation}
  \rho \left(\frac{\partial \theta}{\partial \tau} + u_{x} \frac{\partial \theta}{\partial x} + u_{y} \frac{\partial \theta}{\partial y} \right) =  \nabla^2 \theta + \Omega -b\theta,
\label{e.energy}
\end{equation}
can be solved for the dimensionless temperature variable distribution $\theta = (T -T_u)/(T_a-T_u)$ with constant transport coefficients $\rho D_T = \text{const}$. The parameter ${b= 2h\lambda_w/((h/\delta_T)^2 h_w \lambda_u)}$,  assumed to be of order unity, {\color{black}characterizes the off-plane conductive heat losses through the plates and represents the relative importance of the off-plane conductive heat losses through the plates. This parameter, with clear physical meaning, is obtained using a rigorous asymptotic expansions\citep{fernandez2018analysis, Martinez2019} in the limit $\delta_i << h$. } 
It involves, in its definition, the wall thickness $h_w$, and the walls $\lambda_w$ and burned gas $\lambda_u$ conductivities. The off-plane conductive heat losses term $b \theta$, that appears in the right hand side of equation~\eqref{e.energy}

The conservation of reactant species reads
\begin{equation}
  \rho \left( \frac{\partial Y}{\partial \tau} + u_{x} \frac{\partial Y}{\partial x} + u_{y} \frac{\partial Y}{\partial y} \right) = \frac{1}{Le} \nabla^2 Y - \Omega.
\label{e.species}
\end{equation}

{\color{black} 
Preferential diffusion is included in the model through $Le$, which, in the case of ultralean hydrogen flames is very well characterized \citep{smooke1991reduced} by a value $Le=0.3$. Notice the essential effect of preferential diffusion is the enrichment of the mixture because of curvature effects, which produces superadiabatic temperatures that allow the flame to exist below the usual flammability limit of planar flames\citep{sanchez2014recent} $\phi_l \sim 0.25$. This effect is the primary one and we have not included the Soret effect, which is anticipated to produce relatively minor quantitative corrections\citep{fernandez2011structure, fernandez2023minimum}.}
The system requires the equation of state $\rho\left( 1+q \theta \right) = 1$ and the simplified continuity equation in terms of pressure, $ {\nabla^2 p = -q  ( \triangle \theta + \Omega - b\theta)}$. The latter is used to compute the velocity components as a function of the pressure field from the momentum equation in the viscous limit
\begin{equation}
    {u_{x} = - (\partial p/\partial x)} \quad {\rm and}\quad   u_{y} = - (\partial p/\partial y). \label{eq:mom}
\end{equation}
The reaction rate is modeled using chemical diffusion, that represents adequately the behavior of lean flames while reducing the computational cost of fully detailed kinetics \citep{chung2020structure},
\begin{equation}
  \Omega = \rho^{2}\beta^{2} \left( 1+q \right)^{2}\frac{Y}{2 Le}\exp\left( \frac{\beta\left( \theta -1 \right)}{1 + \frac{q}{1+q}\left( \theta -1 \right)} \right),
\end{equation}
which is a one-step irreversible Arrhenius model \textcolor{black}{which is well characterized for the cases under study by a moderately large Zeldovich number ${\beta = E_a(T_a - T_u)/(R T_{a}^{2})} \approx 10$}, being $E_a/R\simeq 20\times 10^3$~K the characteristic value of global activation temperature for lean hydrogen flames \citep{fernandez2009one}.

The time-evolving simulations lead to two distinct steady propagation states under the same set of numerical parameters, encompassing chemical properties (equivalence ratio $\phi$, activation energy $E_a$, heat release $q$) and boundary conditions (constrained flow and conductive heat losses $b$). The steady solution that finally emerges as $\tau \gg 1$ depends only on the initial conditions set at $\tau=0$. Given appropriate initial profiles, $\theta(x,y;\tau = 0)$ and $Y(x,y;\tau=0)$, the transient process is integrated in time with a third-order Runge-Kutta algorithm of time step $\Delta\tau = 10^{-6}$, limited by reaction rate terms. The interested reader is referred to Domínguez-González et al.\cite{dominguez2024numerical} for additional details.

The characteristic final steady-state  temperature fields of the circular and double-cell canonical flames are illustrated in figures ~\ref{fig:simul}(\textbf{a}) and \ref{fig:simul}(\textbf{b}) respectively. In addition, the reacting front is depicted using white isolines at 15\% and 90\% of the maximum value of the reaction rate $\Omega_{\rm max}$ in each case.

Figures~\ref{fig:simul}(\textbf{c}) and~\ref{fig:simul}(\textbf{d}) display the distribution of reactant concentration $Y$ alongside the flow streamlines (orange). These results indicate more intense reaction rates in double-cell flame fronts, resulting in a propagation speed $U_{L_2}$ approximately double that of the circular counterpart $U_{L_1}$. In the latter, the lower velocity permits higher diffusion of reactants, sustaining the chemical reaction in the rear flame region. Conversely, higher velocities in two-cell flames impede fuel from reaching the rear of the reacting surface through mass diffusion, leading to an elongated shape with a distinct hydrogen-depleted zone behind.
Most important, the emergence of each fixed-point structure (within a common region of the parametric space in the ultra-lean limit) depends on the initial conditions set for the numerical simulations. This suggests that the ignition-like transient is relevant to the development of either the weaker or the stronger solution, as it will be discussed in the following experimental measurements.

\section{Experimental setup and processing}
The simplifications in the formulation of the problem introduced in the previous section to reduce the computational cost (quasi-two-dimensional problem, viscous flow, fast conductive heat losses or reduced 1-step chemical model) raise questions about the actual existence of the two types of flame structures in real devices. To investigate their simultaneous appearance, we designed and constructed the horizontal setup shown in figure~\ref{fig:expsetup} that eliminates the effect of gravity in the flame propagation direction. It consists of two 25~mm-thick methacrylate parallel plates of dimensions $L = 950$~mm and $W = 220$~mm, separated by a gap $h \ll W \sim L$. The gap can be modified in the range $2 \le h \le 6$~mm to promote relative heat losses through the walls and prevent autoinduced acoustic pressure waves from interacting with the reactive front\citep{veiga2019experimental}.

The chamber is filled with a H$_2$–air mixture prepared before injection using two mass flow controllers VÖTGLIN GSC red-y smart series. The chemical heat release and flame thickness is adjusted by carefully tweaking the equivalence ratio $\phi$ of the mixture. The uncertainty in the  initial mixture composition is determined by propagation of errors. The maximum relative deviation is below 2.3\%, which corresponds to ${\Delta \phi = \pm 0.005}$. In order to further reduce the uncertainty, we repeated the experiments using a second set of mass flow controllers, obtaining the same flame behavior in all cases tested. As reported by the manufacturer in their calibration reports, the highest deviation is below 0.12\%, which is well below the calculated uncertainty obtained through propagation of errors. The details of these calculations are included in Appendix A.

Once the chamber volume is filled with the desired reactants, we ignite the mixture at the open end using a three-pin spark plug. The spark-plug is activated using a 12~V battery and a $50$~Hz ($75$~Hz charge cycle) square signal. It should be noted that the electric-arc ignition is not a perfectly reproducible initiation as it involves a stochastic process that slightly changes the details of how the energy is transferred to the gas. To minimize this effect, each experiment was repeated at least three times to confirm a high level of repeatability, consistent with the precision of the controllers (0.5\% of full scale).

\begin{figure}
    \centering
    \includegraphics[width=1\linewidth]{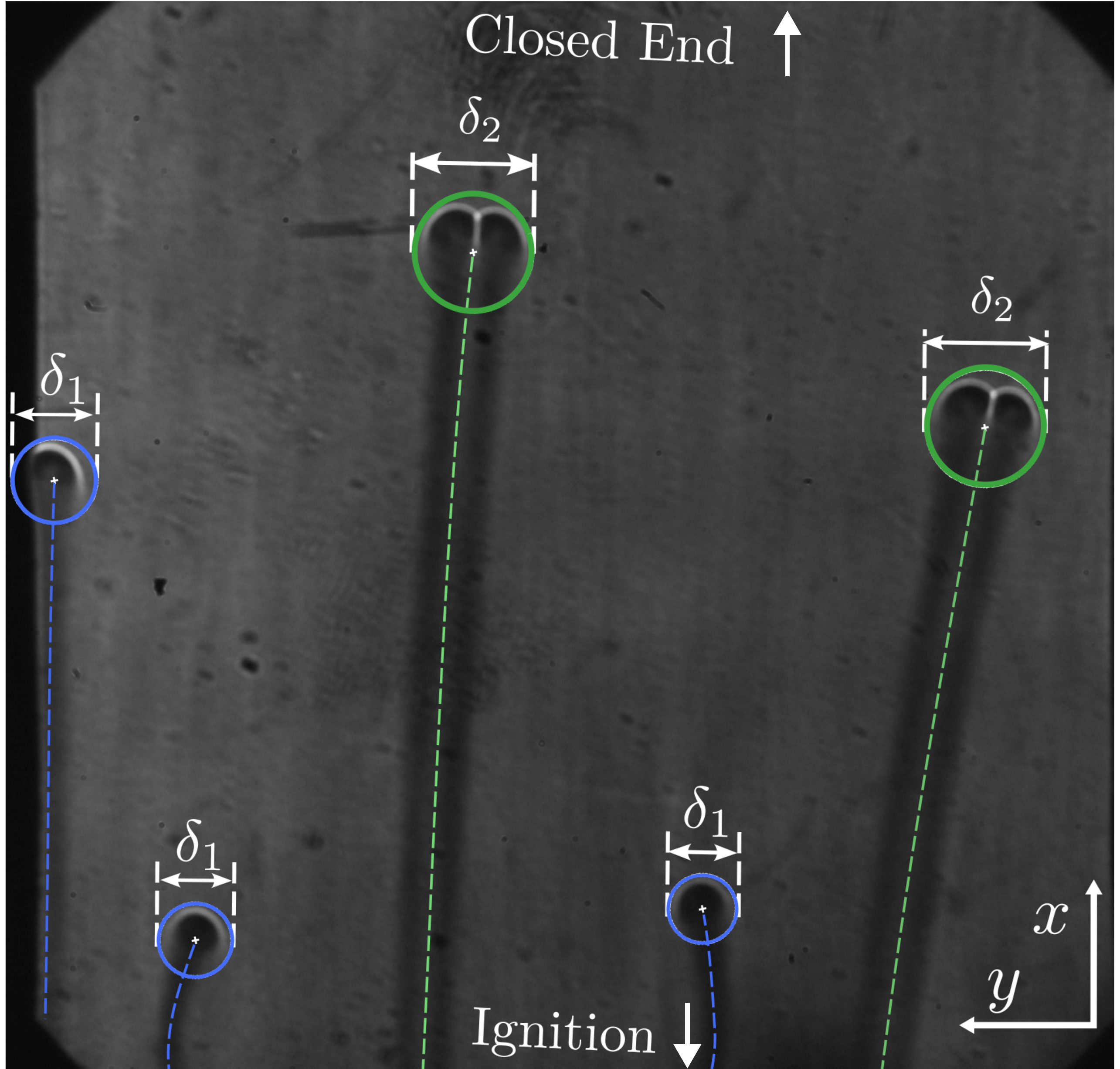}
    \caption{Automatic structure detection over the Schlieren imaging frames. Flame sizes (green and blue circles) and trajectories (dashed lines) are obtained for each kernel through an in-house tracking algorithm.}
    \label{fig:postprocess}
\end{figure}

After ignition, the reaction fronts propagate inside the chamber, which is visually accessible through both plates. The reactive front is visualized through density variations using a Schlieren setup that places both the high-speed camera and a light source perpendicular to the plates, as shown in figure~\ref{fig:expsetup}. A 30 cm-diameter concave mirror is placed below the experiment to reflect the light coming from a $150$~W light source (Fiber-Lite LED Fiber Optic Illuminator) aligned with the camera lens axis. The reflected light beam is intercepted with a razor blade at a distance twice the focal length of the mirror ($2.5$ m). Depending on the particular case, the experiment is recorded in the range of 800 and 1000~fps using a high-speed camera (Memrecam HX-3).
Characteristic experimental images are shown in figure~\ref{fig:postprocess}. The long-lasting darkened trails are caused by condensation of water vapor of the product gases on the surface of the plates \citep{bregeon1978near,kuznetsov2019experiments}, while the hotter flame front can be detected in the picture as a clearer notch. The overall light-gray area illustrates the unburned mixture region. 

The propagating isolated fronts are automatically detected and tracked with an in-house python code using the Open Source Computer Vision Library (OpenCV), that provides contour detection at each frame based on the lighter pixels of the flame. The reactive front is identified by thresholding the images to discriminate between burnt and unburned regions. This algorithm detects each front and identifies every same structure along consecutive time steps, which in turn enables the description of divisions and formation of new elements.  Frame-by-frame post-processing consists on a series of algebraic operations (background subtraction, addition, threshold selection and binarization), which eliminates the need to correct brightness, contrast or defects between experiments. 

Figure~\ref{fig:postprocess} illustrates a sample detection and tracking of an experimental run. The detection of the contour of each individual kernel is used to compute the circular region of minimum radius $\delta_i$ that completely encloses the reactive front at each frame $k$. The coordinates $\vec{x}_i^{\;k}$ of the center of the circle are used to delineate the path followed by the reactive front. Displacement $s=|\vec{x}_i^{\;k}-\vec{x}_i^{\;k-1}|$ and velocity ${U_{L_i}^k = |\vec{x}_i^{\;k}-\vec{x}_i^{\;k-1}| /\Delta t} $ are calculated using backward differentiation with images separated by 70 frames. The estimated uncertainty of the velocity is calculated as $\Delta U_L/U_L = \Delta s/s + \Delta t / t = 0.02$.

\begin{figure*}[t!]
\centering
\includegraphics[width=0.32\textwidth]{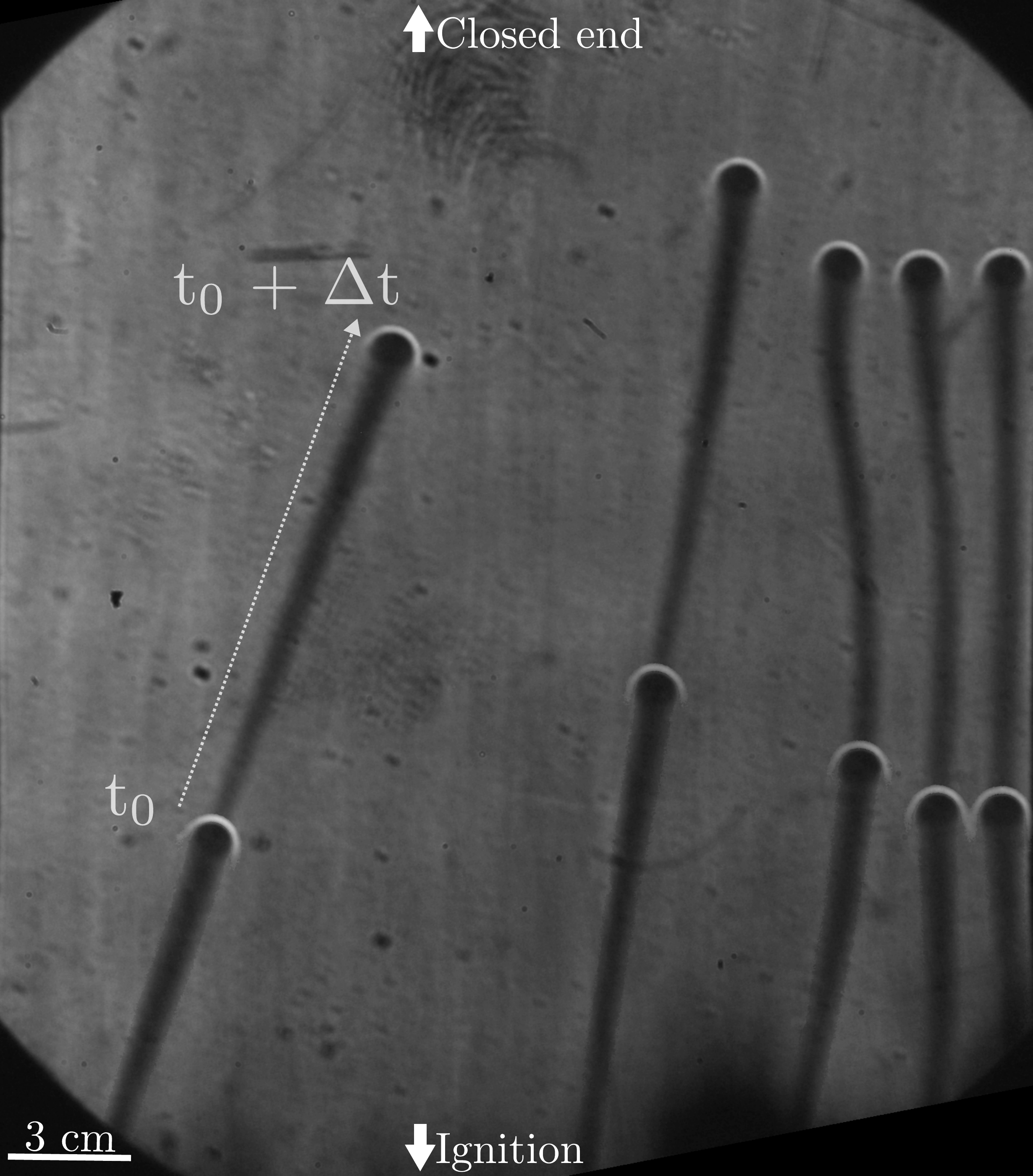}
\includegraphics[width=0.32\textwidth]{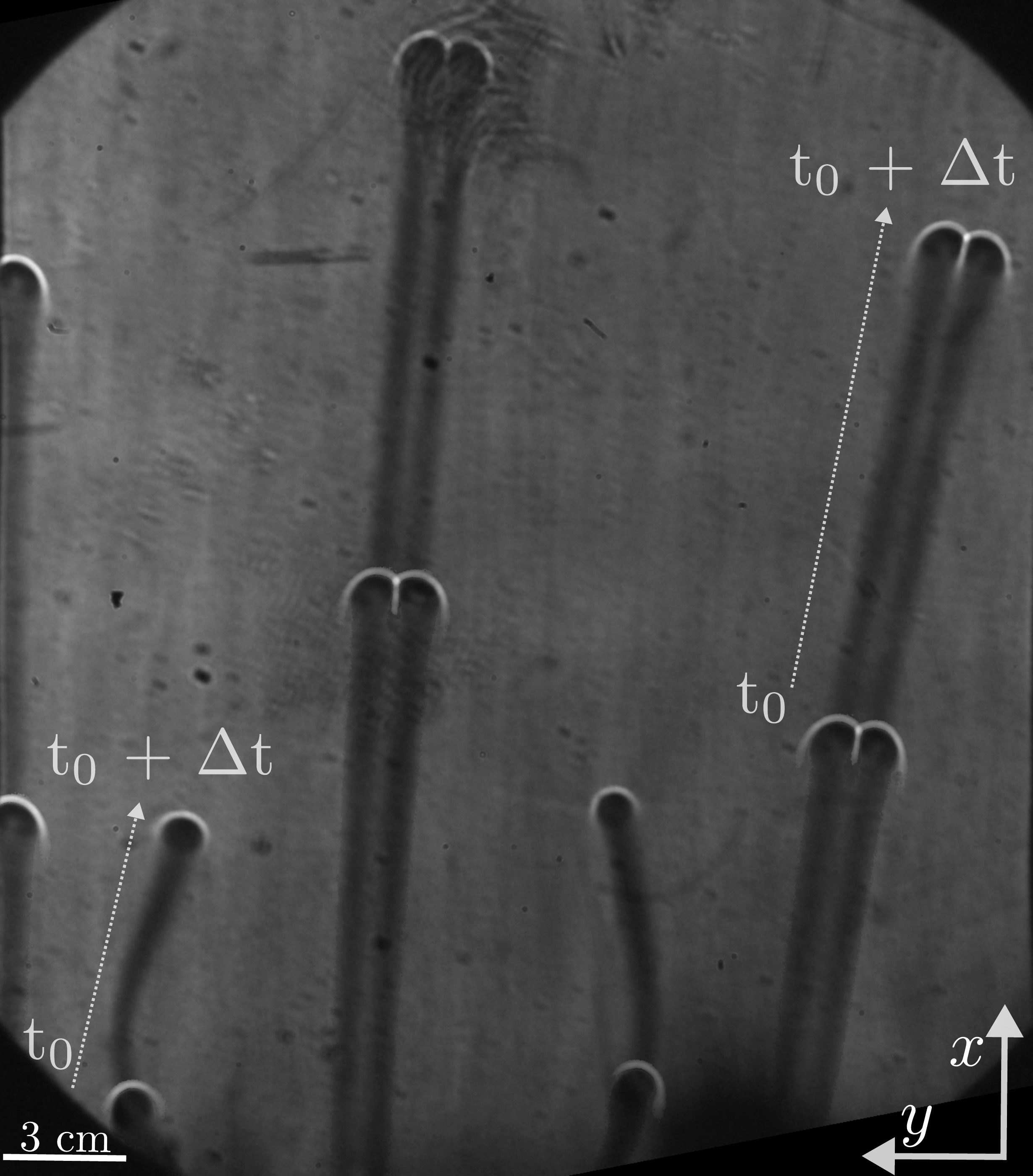}
\includegraphics[width=0.32\textwidth]{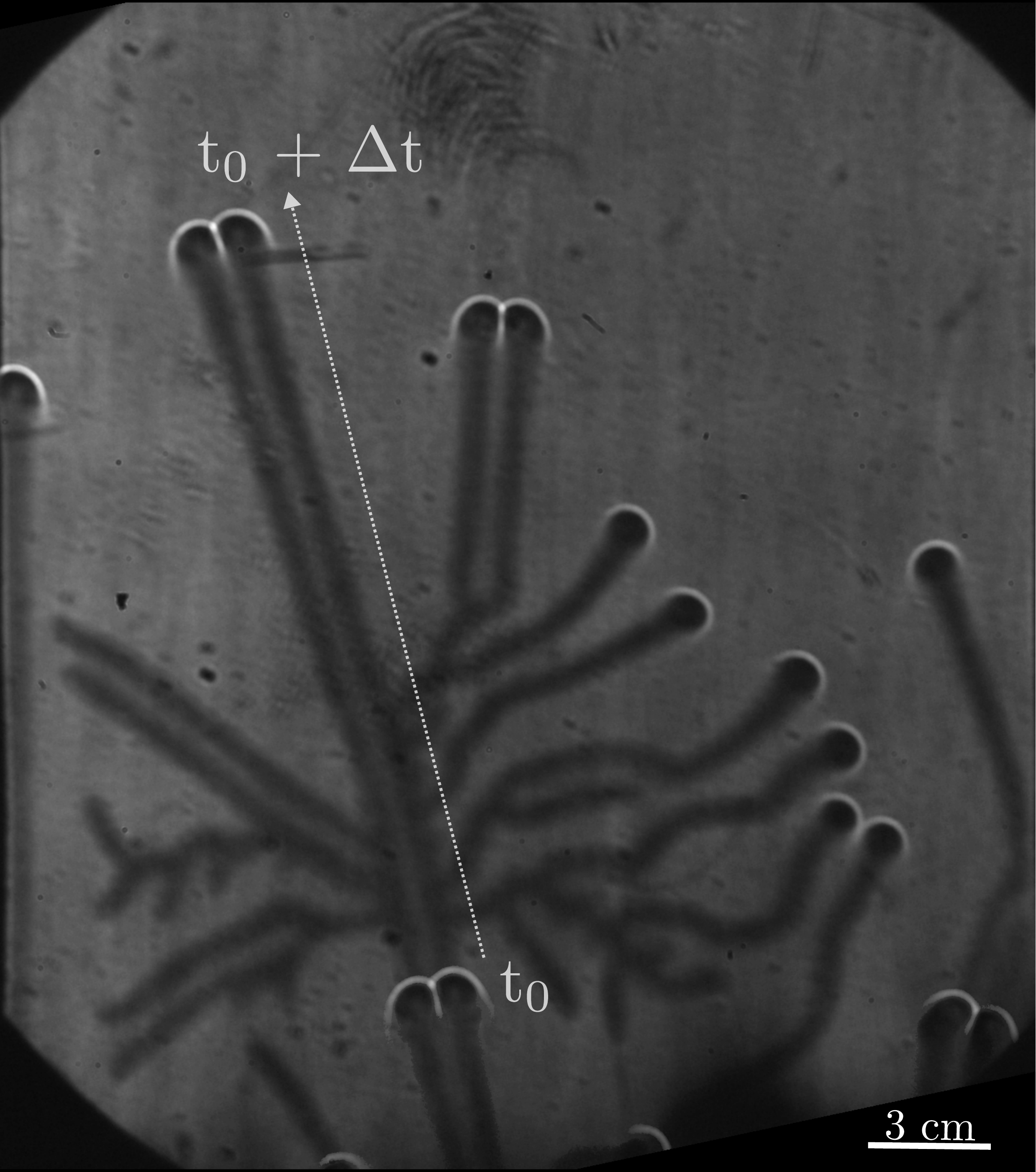}
\caption{Two-frame composition for propagation in channels with $h=4$~mm. Isolated circular flames with \%vol.~H$_2$~$= 7.95$ ($\phi=0.205$), separated a time $\Delta t=2$~s (left). Coexistence of circular and double-cell flames at \%vol.~H$_2$~$= 7.97$ ($\phi=0.206$), $\Delta t=1$~s (center). Splitting fronts with \%vol.~H$_2$~$= 8.01$ ($\phi=0.207$), $\Delta t=1.5$~s (right) (Multimedia available online).
}
\label{fig:one-two-heads}
\end{figure*}

\section{Experimental results}

The following tests have been instrumental to find the simultaneous propagation of two stable flame structures experimentally, confirming the unique bi-stability predicted by the unsteady numerical simulations (figure~\ref{fig:simul}) and providing additional data to map the behavior of lean hydrogen flames. First, in figure~\ref{fig:one-two-heads} (Multimedia available online) we show a composition of experimental images for increasing values of hydrogen concentration with $h=4$~mm. Each panel shows an overlay of two frames, separated a time span $\Delta t$ (case dependent) to show the evolution of the premixed flames.

The left panel depicts a group of five isolated circular flames propagating steadily in nearly straight lines for \%vol.~H$_2$~$= 7.95$ (equivalence ratio $\phi=0.205$).   The time lapse between frames is $\Delta t = 2$~s, and renders equivalent propagation distances of every isolated flame.

\begin{figure*}
\centering
\includegraphics[width=0.95\textwidth]{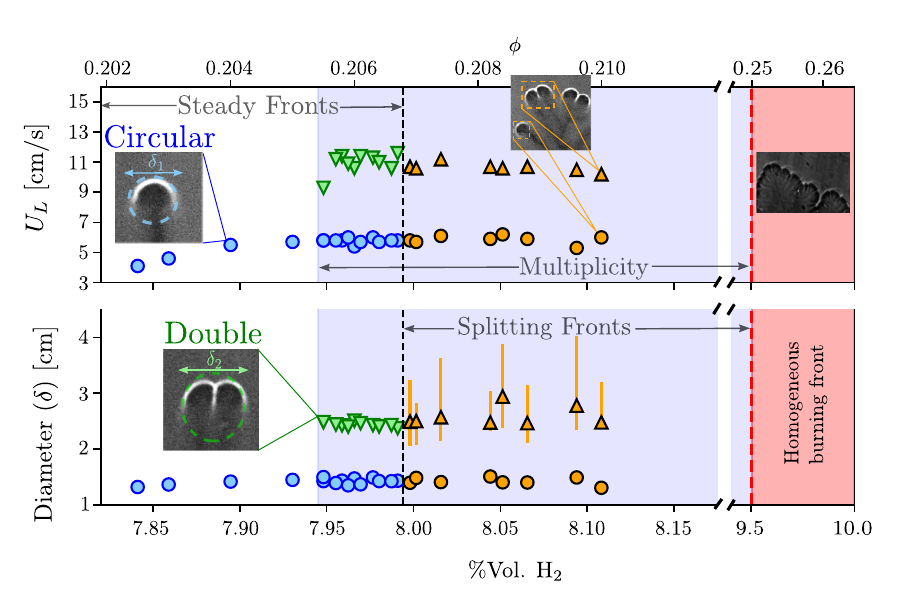}
\caption{Parametric range of multiple stable solutions for $h=4$ mm. Steady propagation velocity $U_{L_i}$ (top) and size $\delta_i$ of the reaction front (bottom) of circular $i=1$ (circles) and double-cell $i=2$ (triangles) flames. Circular flames are always stable while double-cell flames become unstable when $\phi>0.2065$.}
\label{fig:bifurcation}
\end{figure*}

The first evidence of bi-stability emerges spontaneously for a concentration of \%vol.~H$_2$~$= 7.97$ ($\phi=0.206$), shown in the central panel of figure~\ref{fig:one-two-heads}, where we identified both steady circular and double-cell flames simultaneously. They are originated at a single ignition event and travel steadily along the chamber burning a small fraction of the total fuel available. The two-frame overlay (time lapse $\Delta t= 1$~s) shows that double-cell flames move, roughly, two times faster than their circular counterpart. Notice that this double-cell flame appears as a single structure, with a narrow extinction gap marked at their center by the separation of the two parallel condensation trails. This configuration is in clear contrast with the unstable adjacent 2D flame structures found numerically by Grcar\cite{grcar2009new}, in which two isolated flame cells are separated by a gap of unburned gas of roughly the same size as themselves.

Finally, double-cell flames become unsteady for concentrations above \%vol.~H$_2$~$=8.00$. The rightmost panel of figure~\ref{fig:one-two-heads} shows the flame front splitting into  multiple circular and double-cell flames that form an angle with the primary flame displacement. The average flame propagation speed remains constant, but the size of the reactive front oscillates before it eventually divides into more circular and double-cell flames. The frequency of double-cell flame bursting increases with increasing H$_2$ concentration, leading to the continuous breakup of the reactive front when \%vol.~H$_2>9.5$ ($\phi=0.25$).

The collection of experimental results for a channel height of $h=4$~mm is summarized in figure~\ref{fig:bifurcation} as bifurcation diagrams for flame speed $U_{L_i}$ and size of the reactive front $\delta_i$ versus fuel concentration $\phi$. The critical values of $\phi$ can be drawn to determine the flame properties in the various regimes identified above. First, at ultra-lean propagation conditions \%vol.~H$_2<7.94$, only circular flames exist, with characteristic diameters of ${\delta_1 = 1.2}$~cm and steady propagation velocities that range between $U_{L_1}=4$ and $6$~cm/s. This behavior is that of the left panel in figure~\ref{fig:one-two-heads}. Above \%vol.~H$_2$~$=7.95$ ($\phi=0.205$), the system bifurcates into the multiplicity regime causing the simultaneous appearance of two stable configurations, as shown in figure~\ref{fig:one-two-heads}, center. Note that the size and speed of the circular fronts do not involve substantial changes compared to the previous regime. Nevertheless, double-cell fronts exhibit a size $\delta_2=2.5$~cm and propagation speed $U_{L_2} = 11$~cm/s twice as large as their circular counterparts. Furthermore, the unsteady splitting regime is found for greater hydrogen content than \%vol.~H$_2 \simeq 8.00$ ($\phi=0.207$). Above this critical value, the size of the double-cell front undergoes another transition from a steady state to a cyclic oscillation branch in which the leading flame grows and shrinks periodically before splitting to form circular flames seeded perpendicularly to the reactive path, see right panel in figure~\ref{fig:one-two-heads}. 
The amplitude of these oscillations is small, and the limited resolution of the images does not allow for their accurate quantification except right before splitting takes place. This is precisely what we represent using the vertical lines to denote the maximum amplitude of these oscillations in the lower panel of figure~\ref{fig:bifurcation}. Both velocity and size of each type of flame remain constant during the whole concentration range tested in the experiments, including the region associated to double-cell cyclic oscillations.

Finally, equivalence ratios even greater than \%vol.~H$_2 \simeq 8.12$, produce enough heat release to avoid partial extinction of the front and formation of isolated structures. In that case, a continuous corrugated and unstable flame front is formed, which burns the whole volume of available reactive mixture.

\begin{figure*}
	\centering
	\includegraphics[width=0.32\textwidth]{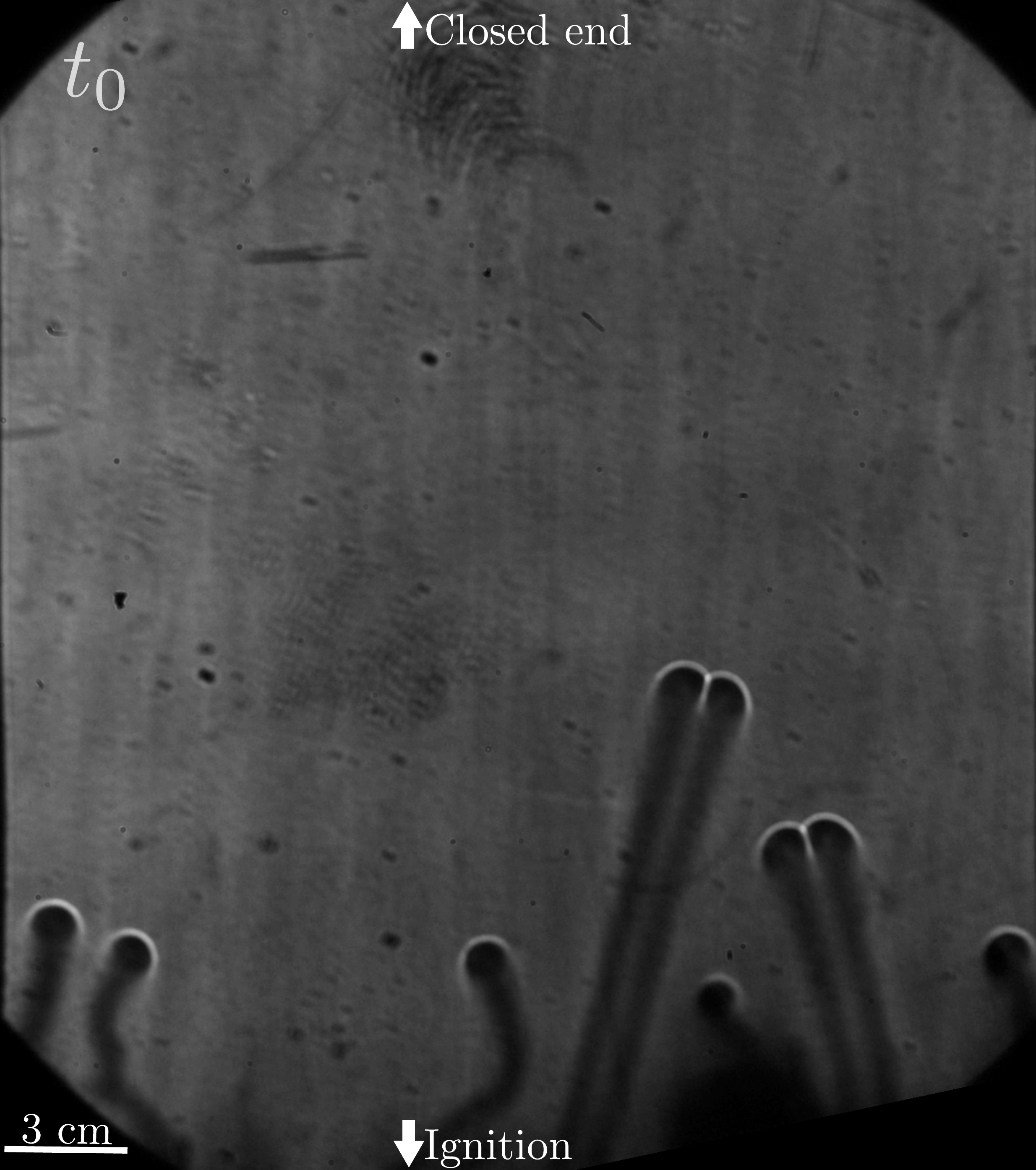}
	\includegraphics[width=0.32\textwidth]{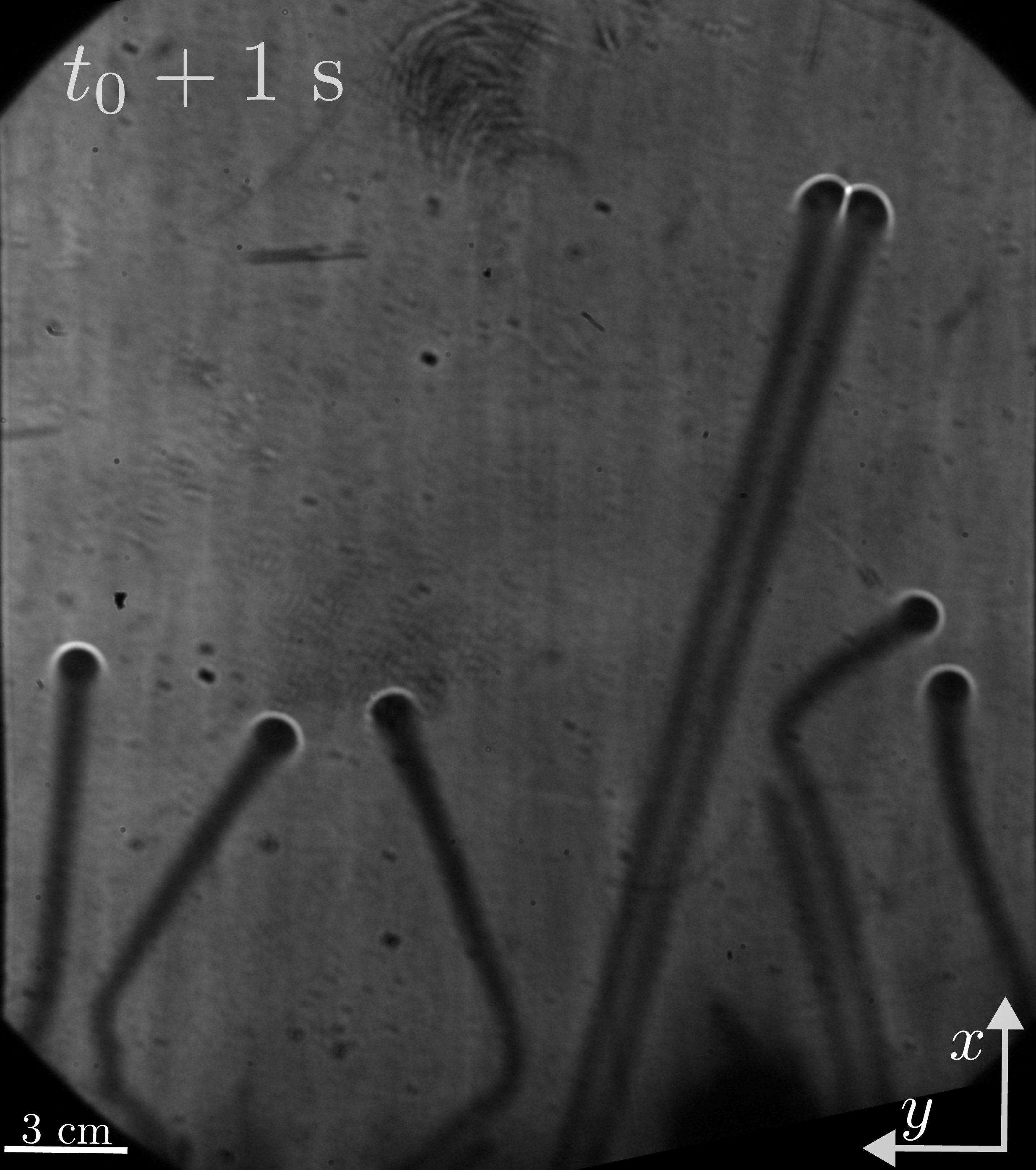}
	\includegraphics[width=0.32\textwidth]{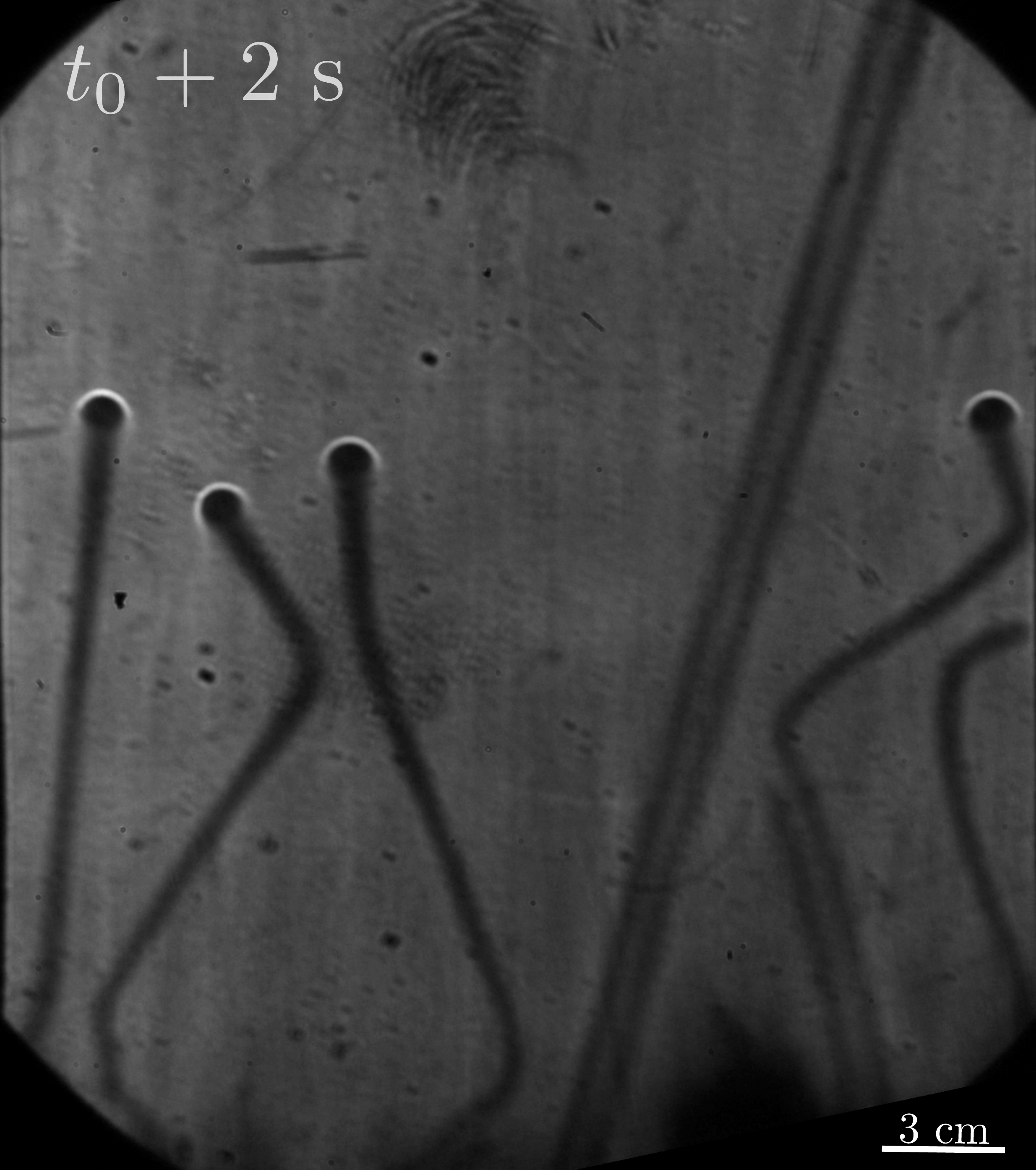}
	\caption{Time sequence for the simultaneous propagation of circular and double-cell flames at \%vol.~H$_2=7.98$ ($\phi=0.206$). Interactions between kernels and with the lateral walls of the chamber can be observed (Multimedia available online). 
	}
	\label{fig:one-two-heads_sm}
\end{figure*}

Although these changes in behavior and critical limits are obtained consistently for a specific separation of the plates, the relative importance of heat losses was also modified by changing the gap size in the range $2 \le h \le 6$~mm. The behavior of the flame remains qualitatively similar to what has been described above for $h=4$~mm, with both circular and/or double-cell flames emerging when the equivalence ratio becomes sufficiently small ${\phi<[0.181, 0.198, 0.210, 0.240, 0.277]}$ for ${h=[6, 5, 4, 3, 2]}$~mm respectively.

In contrast, stable circular and double-cell flames coexist exclusively for $h=3$ and $h=4$ mm gap sizes. In smaller ($h=2$ mm) or larger gaps ($h=5$ and $6$ mm), the coexistence of solutions is restricted to stable circular flames alongside unstable double-cell flames that split periodically. This variation in flame behavior indicates that the heat loss-to-production balance, $\Delta = q_k/q_f$, becomes either too small or two large to enable the steady state propagation behavior of both kinds of kernels. In fact, the splitting regime emerges when this balance is broken. The thermal energy source increases with higher hydrogen content in the mixture (greater $\phi$), elevating the flame and exhaust gas temperatures. When heat losses can no longer balance heat production, the flame grows and its tail starts oscillating, initiating the front-splitting phenomenon shown in figure~\ref{fig:one-two-heads} (right).

\begin{figure}
    \centering
    \includegraphics[width=0.5\linewidth]{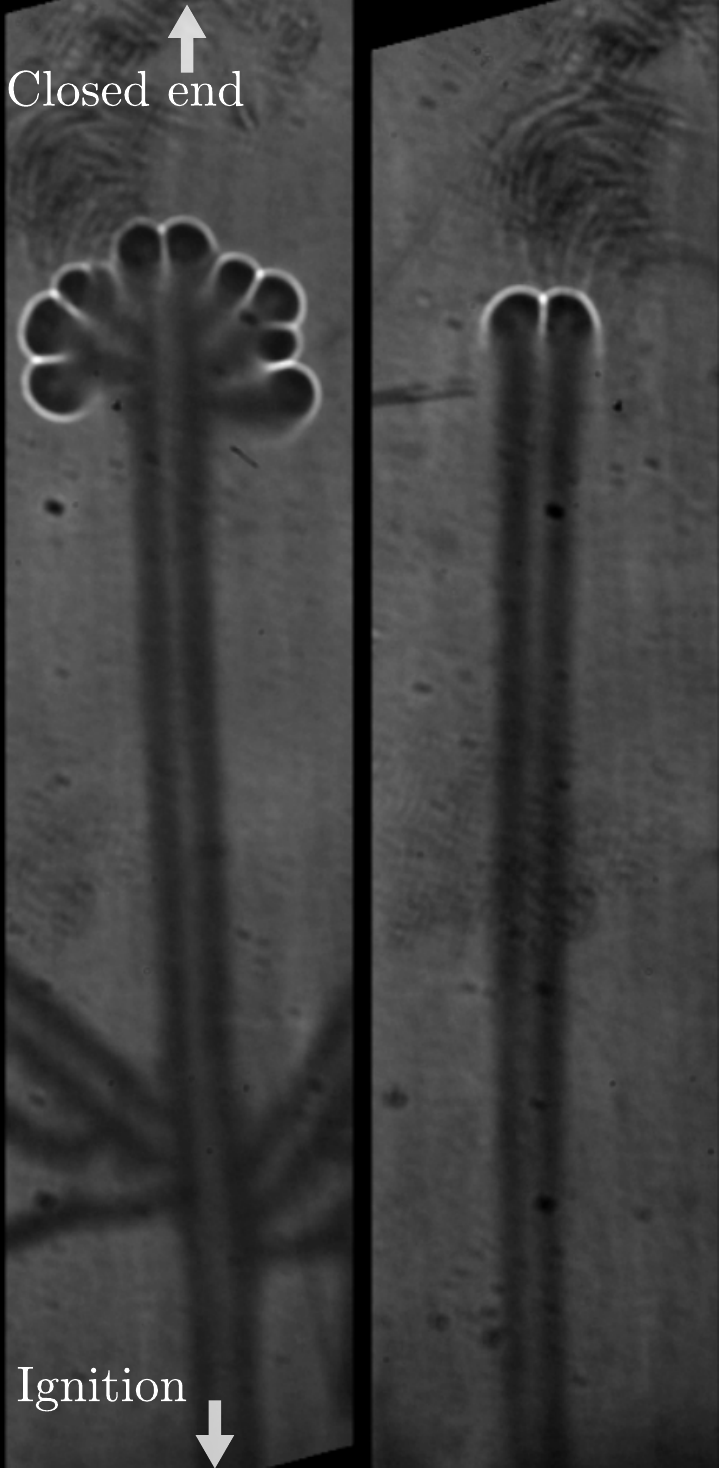}
    \caption{Comparison between the propagation of a stable ($\phi=0.206$) and an unstable double-cell flame ($\phi=0.208$). (Multimedia available online)}
    \label{fig:stable-unstable}
\end{figure}

Additionally, the propagation of a stable ($\phi=0.206$) and an unstable oscillating ($\phi=0.208$) double-cell flame is compared in figure \ref{fig:stable-unstable} (Multimedia available online) In the unstable case, the flame oscillates at approximately 26.6~Hz before splitting, generating multiple circular and double-cell flames. Following this event, the double-cell flame continues its propagation before resuming flickering at the same frequency. In spite of the oscillations, the propagation speed remains unaltered. 
Furthermore, different events arise when interacting with the bounding lateral walls or between kernels. An illustrative example is given in figure~\ref{fig:one-two-heads_sm}, including a sequence of images for \%vol.~H$_2=7.98$ in which several circular flames coexist with a double-cell flame that propagates steadily. These pictures illustrate the mutual interaction between the reactive fronts, changing the propagation direction but leaving the modulus of the propagation velocity unchanged. Collisions with the sides of the chamber are also observed in the footage, with the only consequence being a change of propagation direction. Contrarily, the change in the morphology of double-cell flames is seen in the middle panel of figure~\ref{fig:one-two-heads_sm}, when one of them approaches the hydrogen-deprived trail left by other reactive front. Under this scenario, the double-cell structure becomes a circular flame that continues its steady burning at a lower rate. These observations suggest that the smaller circular flames may exhibit greater stability to perturbations compared to the double flame structure.

\begin{figure*}
	\centering
	\includegraphics[width=\textwidth]{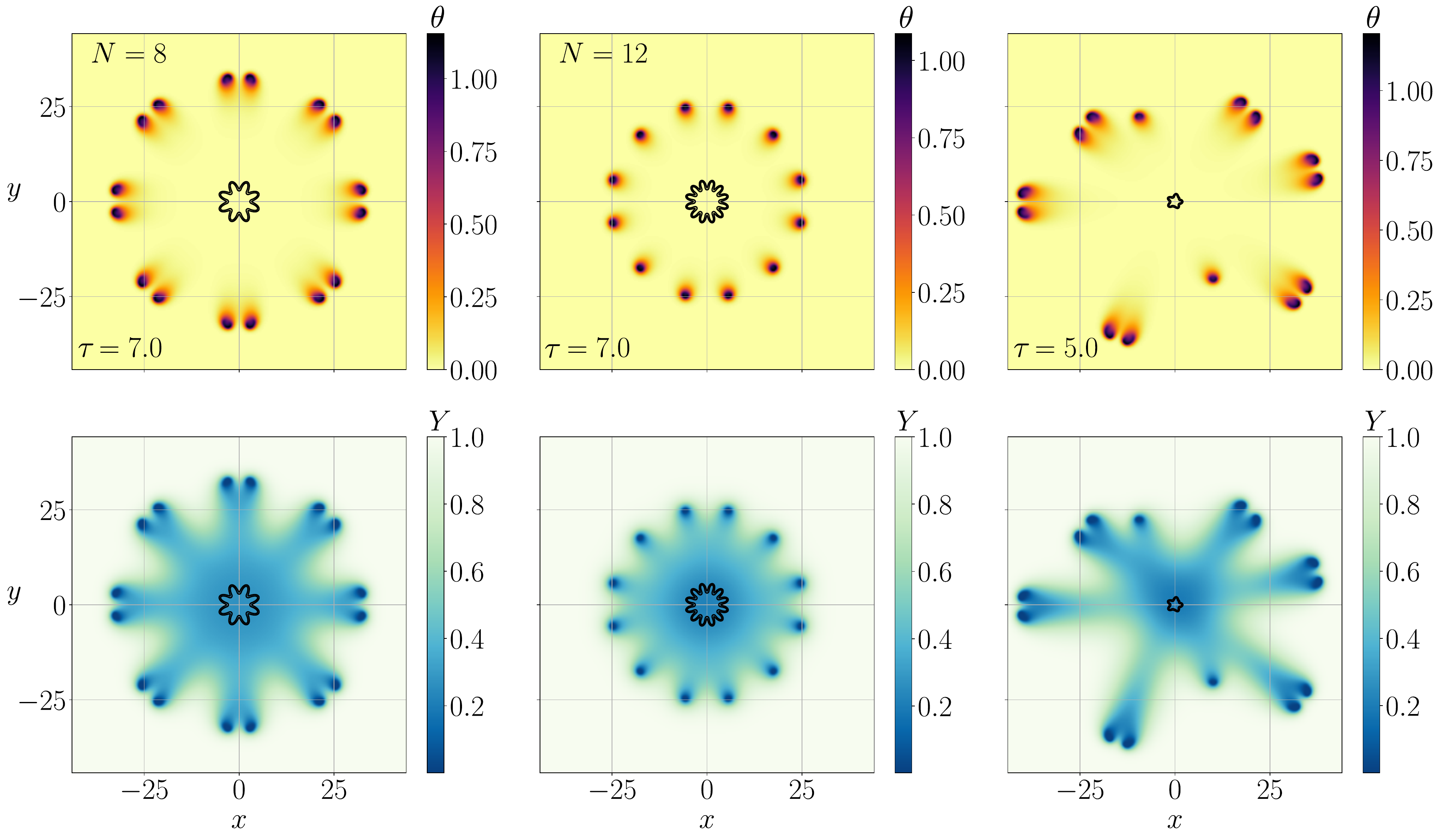}
	\caption{Numerical simulations for different initiation conditions with $N=8$ (left), $N=12$ (center) and irregular composition of modes $N_i = [2, 5, 10, 15, 20, 25, 30] $ (right) distributions. Dimensionless temperature profiles at selected instants are shown (top) together with the reactive species field (bottom) (Multimedia available online). 
	}
	\label{fig:num_initiation}
\end{figure*}

\section{Irregular ignition transient}\label{sec:ignition}

It has been shown that for particular combinations of heat release and relative heat losses (equivalence ratio and channel height), multiple stable solutions may arise. Slight variations in the concentration field, minor fluctuations in ignition spark energy, or subtle heat transfer imbalances between the flame and the plates could explain why a different number of circular or double-cell flames emerged in each experimental run.
As a matter of fact, although both canonical solutions are feasible under these common parametric-space conditions, some symmetry-breaking effects during ignition must determine the actual distribution and propagation of various types of kernels.

To test this hypothesis, we carried out an additional parametric study with several numerical simulations modifying their initial conditions. The initiation profiles involve a hot spot with radial decay, together with consistent species depletion to avoid overshooting of the reaction rate, which are perturbed with $N$ lobes and random noise
\begin{equation}
\theta(\tau=0)=\theta_0 \exp \left(- \frac{r^2}{10} \left[1+ \sum_{i=1}^{m} \frac{\cos (N_i \theta+ 2\pi\varepsilon_i)}{2}\right]\right),
\end{equation}
\begin{equation}
Y(\tau=0)=1-\exp \left(- \frac{r^2}{10} \left[1+ \sum_{i=1}^{m} \frac{\cos (N_i \theta+ 2\pi\varepsilon_i)}{2}\right]\right),
\end{equation}
where $r^2= x^2+ y^2$ is the dimensionless radial coordinate, $m$ is the number of different modes included in the initiation profile, $N_i$ is the wave-number and $\varepsilon_i \in[0,1]$ is a computer-generated random shift of each mode. In axi-symmetric ignition cases $\varepsilon=0$ to generate a single mode $N$. 

Figure~\ref{fig:num_initiation} (Multimedia available online) shows numerical solutions of temperature (top) and species (bottom) under a combination of dimensionless heat release $q=5$ and conductive heat-loss coefficient $b = 1$, selected to ensure bi-stable solutions~\citep{dominguez2022stable}. Moreover, the same peak temperature $\theta_0 = 1.8$ is selected in all cases with different initial shape distribution, determined by $N=8$ modes (left panels), $N=12$ modes (center panels) and a superposition of seven different modes $N_i=[2, 5, 10, 15, 20, 25, 30]$ with random shift. Different regular initiations at $\tau \ll 1$ (shaped in black isocontours in figure~\ref{fig:num_initiation}) induce the formation of either double ($N=8$) or circular ($N=12$) flames, owing to the transient growth and splitting dynamics of the propagating front. However, non-regular initiation cases evolve into corrugated shapes that do not show azimuthal periodicity, nor axisymmetric solutions with only one kind of kernels. Therein, two different families of flames arise, of different sizes and which propagate at different velocities. Therefore, we expect that minor irregularities in the ignition process suffices to alter the transient history of the front and give rise to both stable-propagating structures, provided that the balance $q$ and $b$ (or $q_k/q_f$) is adequate for multiplicity. 

To complete this analysis, we include below in figure~\ref{fig:ignition} an experimental visualization of the area near the spark plug right after the ignition. The circular regions at the bottom of the image are the pressure sensor (blue circular region) and the plastic piece that holds the spark plug (gray region). They are both opaque, preventing transverse flame visualization with the Schlieren setup. Shortly after the spark ignition, the front is divided into multiple flames that evolve differently while propagating radially. In this figure we identify two types of reactive fronts propagating in different direction shortly after ignition. After observing the numerical results shown in figure \ref{fig:num_initiation}, we hypothesize that the irregularities during the ignition process produced the necessary variations in the flow field to facilitate the emergence of both single and double flames over the same mixture composition $\phi$ and chamber height $h$. The control of the ignition details to produce perfectly symmetrical initiation cannot be addressed with the actual experimental set-up and should be studied with additional techniques and ignition devices. 

\begin{figure}
	\centering
	\includegraphics[width=\linewidth]{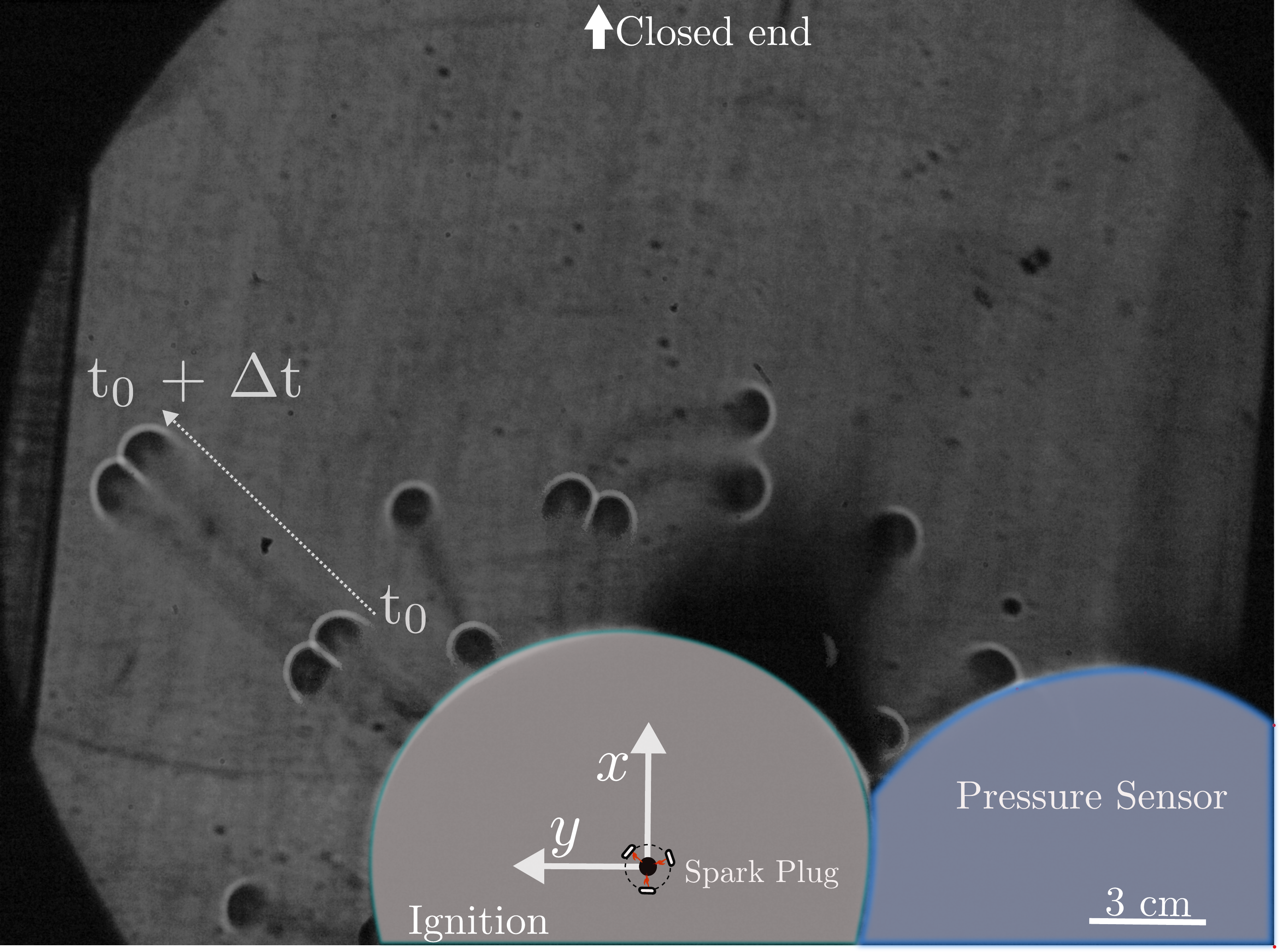}
	\caption{Flame structures generated right after the ignition event. Single and double flames can be observed outside the opaque piece mounting the spark plug (gray region) and the pressure sensor (blue region). The time difference between the two images is $\Delta t=0.5$s.
	}
	\label{fig:ignition}
\end{figure}

\section{Conclusions}
In non-linear systems, the presence of multiple solutions typically arises from the equilibrium of competing physical mechanisms. In combustion theory, solution uniqueness fades away in one-dimensional planar flames due to factors like competing non-linear chemical reactions, molecular diffusion, heat losses, or phase-change processes \citep{clavin1987multiplicity, bonnet1995non}. Fixed-point multiplicity can also be observed in higher-dimensional simulations or experimental combustion setups, often linked to hysteresis loops in dynamical systems, such as thermoacoustic coupling \citep{etikyala2017change}, flame edges \citep{daou2003effect}, droplet combustion \citep{carpio2019hysteresis}, and cool flames \citep{reuter2016experimental}. In these previous examples, solution multiplicity is unveiled by slowly modifying the controlling parameter. Typically, the experiment or simulation starts below a critical value when the system is at a stable steady state. As the value of this parameter is increased, the systems bifurcates and either starts oscillating or evolves to achieve a different steady state. If the experiment is repeated reducing the value of the parameter from a larger initial value, the systems evolves following a different path in which the unsteady solution is not observed or the steady solution is different. Farther reduction of this parameter to reach the critical value would give a sudden transition to return to the initial fixed point.  Throughout all these examples, the evolution of the system is determined by its initial state and the bi-stability of the system is only observed in the hysteresis region. Outside this region, both solutions were never observed simultaneously.

A key distinction between these examples and our findings lies, precisely, on the absence of hysteresis in our results and the simultaneous observation of two stable solutions. The same behaviour is observed regardless of how the equivalence ratio, our controlling parameter, is varied in the experiments. Each experiment is completely independent from the others, and the evolution of the system is not affected by its initial state. In other words, the simultaneous observation of two stable solutions does not depend on how the equivalence ratio is modified. 

The critical equivalence ratio $\phi \simeq 0.205$ identifies the bifurcation point above which stable circular and double-cell reactive fronts coexist within a single experimental. The double-cell solution losses stability for $\phi > 0.207$, with the flame oscillating at constant frequency before splitting, while the circular reactive front remains stable. This region extends up to $\phi \simeq 0.25$, when the homogeneous burning front solution is recovered. 

Finally, the numerical predictions and experimental verification of this study indicate that the emergence of the two stable configurations is determined by symmetry-breaking details during ignition transients before reaching the final steady state. The importance of  stochasticity during the ignition process to explain the multiplicity of solution found in the experiments has been tested by changing the initial condition in our numerical simulations. The resemblance of the initial stages of the propagation to the experimental results is very good, confirming the relevance of the ignition details to explain the onset of bistability and the subsequent evolution of the reactive front

The differences in morphology and burning speed are the kind of fundamental aspects that need to be understood to ensure a safe deployment of technology for low-emission hydrogen, whose production is expected to reach 20 Mt/year by 2030 \citep{IEA_2023}.  We believe our findings will pave the way for further research to enhance our understanding of the physics of H$_2$ flames, ultimately promoting safer and more efficient utilization of this essential fuel.\\

\begin{acknowledgments}
This work was funded by the Regional Government of Madrid and by MCIN of Spain with funding from European Union NextGenerationEU (PRTR-C17.I1) and by Agencia Estatal de Investigación under projects TED2021-129446B-C41 \& TED2021-129446B-C43 (MICINN/FEDER, UE).

The authors wish to thank the technical knowledge and assistance of David Díaz, Israel Pina and Manuel Santos in the design and construction of the experimental setup. 

Declaration of Interests. The authors report no conflict of interest.
\end{acknowledgments}

\section*{Data Availability Statement}
The data that support the findings of this study are available from the corresponding author upon reasonable request.

\appendix

\section{Error propagation}
To check the accuracy of the measurements, we repeated the experiments using two sets of VÖGTLIN GSC red-y smart series controllers (high-performance and standard controllers), with the highest deviation being below 0.12\%, as reported by the manufacturer in the calibration reports. Within the margin of error, the threshold values of the equivalence ratio separating different flame behaviors were identical with both sets of controllers.
The uncertainty in the measurement of the equivalence ratio was determined using the propagation of errors theory and it represents the maximum deviation expected in the performed experiments. This value was calculated using the following expression
$$\left |\dfrac{\Delta \phi}{\phi}\right | = \left |\dfrac{\Delta \dot{m}_F}{\dot{m}_F} \right |+ \left | \dfrac{\Delta \dot{m}_a}{\dot{m}_a}\right |,$$ 
with $\Delta \dot{m}_F$ and $\Delta \dot{m}_a$ the accuracy of the hydrogen and air mass flow controllers, respectively. The values of $\Delta \dot{m}_F$ and $\Delta \dot{m}_a$ are provided by the manufacturer, being $\pm0.3$\% of full scale plus  $\pm0.5$\% of reading in the case of high-performance controllers and $\pm1$\% of full scale if a standard controller is used.  

As seen in the derivation above, the uncertainty depends on the specific mass flow rates used in each experiment.
Hence, the values of $\dot{m}_F$ and $\dot{m}_a$ in our experiments were carefully selected to maintain the measurements as accurate as possible. In the most unfavorable case $\phi=0.2$, we chose $\dot m_a=80.0$ g/min, $\dot m_F=0.477$ g/min, $\Delta \dot m_a/ \dot m_a=\pm 0.0121$, $\Delta \dot m_F / \dot m_F= \pm 0.0107$, leading to $\Delta \phi/\phi < 2.3$\%, or $\Delta \phi=\pm0.005$.

\bibliography{scibib}

\end{document}